\documentclass[twocolumn]{emulateapj}

\bibpunct{(}{)}{;}{a}{}{,}

\newcommand{\msun}{\mbox{$M_\odot$}}
\newcommand{\rsun}{\mbox{$R_\odot$}}
\def\be{\begin{eqnarray}}
\def\ee{\end{eqnarray}}
\def\lsim{\mathrel{\rlap{\lower3pt\hbox{\hskip1pt$\sim$}}
     \raise1pt\hbox{$<$}}} %less than or approx. symbol
\def\gsim{\mathrel{\rlap{\lower3pt\hbox{\hskip1pt$\sim$}}
     \raise1pt\hbox{$>$}}} %greater than or approx. symbol

\shorttitle{GRBs \& Hypernova}
\shortauthors{Brown, Lee, Moreno M\'endez}

\begin{document}

\title{GRBs and Hypernova Explosions of Some Galactic Sources}

%\author
%{Gerald E. Brown,$^{1}$ Chang-Hwan Lee$^{2\ast}$ and Enrique Moreno M\'endez$^{1}$\\
%\normalsize{$^{1}$Department of Physics and Astronomy, State University of New York,}\\
%\normalsize{Stony Brook, New York 11794, USA}\\
%\normalsize{$^{2}$Department of Physics, Pusan National University, Busan 609-735, Korea}\\
%\normalsize{$^\ast$To C.-H. Lee correspondence should be addressed; E-mail: clee@pusan.ac.kr.} }

\author{Gerald E. Brown}
\affil{Department of Physics and Astronomy,
               State University of New York, Stony Brook, NY 11794, USA.}
\email{GEB: gbrown@insti.physics.sunysb.edu}

\author{Chang-Hwan Lee}
\affil{Department of Physics, Pusan National University,
              Busan 609-735, Korea.}
\email{CHL: clee@pusan.ac.kr}

\and

\author{Enrique Moreno M\'endez}
\affil{Department of Physics and Astronomy,
               State University of New York, Stony Brook, NY 11794, USA.}

\email{EMM: moreno@grad.physics.sunysb.edu}

%---------------------------------------------------------------------

\begin{abstract}
Knowing the Kerr parameters we can make quantitative calculations of the rotational energy of black holes. We show that Nova Sco (GRO J1655$-$40), Il Lupi (4U 1543$-$47), XTE J1550$-$564 and GS 2023$+$338 are relics of gamma-ray burst (GRB) and Hypernova explosions.  They had more than enough rotational energy to power themselves.  In fact, they had so much energy that they would have
disrupted the accretion disk of the black hole that powered them by the communicated rotational energy, so that the energy delivery was self limiting.  The most important feature in producing high
rotational energy in the binary is low donor (secondary star) mass.

We suggest that V4641 Sgr (XTE J1819$-$254) and GRS 1915$+$105 underwent less energetic explosions; because of their large donor masses. These explosions were one or two orders of magnitude lower
in energy than that of Nova Sco. Cyg X$-$1 (1956$+$350) had an even less energetic explosion, because of an even larger donor mass.

We find that in the evolution of the soft X-ray transient sources the donor (secondary star) is tidally locked with the helium star, which evolved from the giant, as the hydrogen envelope is stripped off in common envelope evolution. The tidal locking is transferred from the helium star to the black hole into which it falls. Depending on the mass of the donor, the black hole can be spun up to the angular momentum necessary to power the GRB and Hypernova explosion. The donor decouples, acting as a passive witness to the explosion which, for the given angular momentum, then proceeds as in the Woosley Collapsar model.

High mass donors which tend to follow from low metallicity give long GRBs because their lower energy can be accepted by the central engine.

\end{abstract}


%--------------------------------------------------------------------
\section{Introduction}\label{}

The hypernova explosions accompanying GRBs are Type I$_{bc}$;
i.e., they show no hydrogen lines and no helium lines.
Arguments have been given that the helium lines would not be seen
even if the helium were present, that helium would have to mix
with $^{56}$Ni if the lines were to be seen, etc.
Thus, hydrogen is not present at the time of the explosion.
As we shall outline, this is the situation in common envelope evolution in
Case C mass transfer. Case C mass transfer means mass transfer after the helium burning of the giant is finished.
For such a case the GRB and hypernova explosion for Nova
Sco (GRO J1655$-$40) was described by \citet{Bro00}.  We can
now reconstruct the explosion for this case, since the Kerr
parameters ($a_{\star}$) of Nova Sco and Il Lupi have been measured
in the Smithsonian-Harvard-MIT observations \citep{Sha06}, with $a_{\star}=0.65-0.75$ and $a_{\star}=0.75-0.85$
respectively. They check against the prediction of \citet{Lee02} (denoted as LBW) who found $a_{\star}=0.8$ for both. From the
$a_{\star}$ we can construct available energies in the
Blandford-Znajek mechanism. We have a simple guiding principle for
the sources considered; namely, that the explosion energy depends
chiefly on the mass of the donor (secondary star), and this is
easily seen if the binaries are evolved in Case C mass transfer,
as we shall show.

In similar vein, the GRBs and hypernova explosions can be
constructed for XTE J1550$-$564 and GS 2023$+$338 (V404 Cygni),
the available rotational energy being nearly the same as in {Nova
Sco}.

\citet{Mor07} also reconstructed the explosions of
GRS 1915$+$105 and V4641 Sgr.  They found the
explosion of Cyg X-1, in agreement with \citet{Mir03},
to be a dark explosion; i.e., orders of magnitude less
explosive than Nova Sco.

\section{Role of Donor Star in Common Envelope Evolution}

Using the relation between the He core mass ($M_{\rm He}$) of a giant after finishing H-core burning and the initial giant mass ($M_{\rm giant}$),
\be
M_{\rm He}=0.08(M_{\rm giant}/\msun)^{1.45}\msun,
\label{eq:MHe}
\ee
LBW found that following
common envelope evolution, \be a_f \simeq
\frac{M_d}{\msun}\left(\frac{M_{\rm giant}}{\msun}\right)^{-0.55}
a_i. \label{eq:af}\ee  Here $a_f$ is the final separation of the
He star which remains from the giant following the strip off of
its H envelope, and $a_i$ is its initial separation, $M_d$ is the
mass of the donor (secondary star).
% and $M_{\rm giant}$ is the mass of the giant.
Noteworthy about eq.~(\ref{eq:af}) is that the main
dependence of the final separation $a_f$ is on the donor mass
$M_d$, only roughly as the square root of $M_{\rm giant}$.

The He star remainder of the giant and the donor are tidally
locked at the end of common envelope evolution (LBW).
The tidal locking ends here, the Kerr parameter of the black hole
being determined by its angular momentum at formation, minus the
decrease from angular momentum spent in powering explosions.

\begin{figure}[ht]
\centerline{\includegraphics[width=\textwidth]{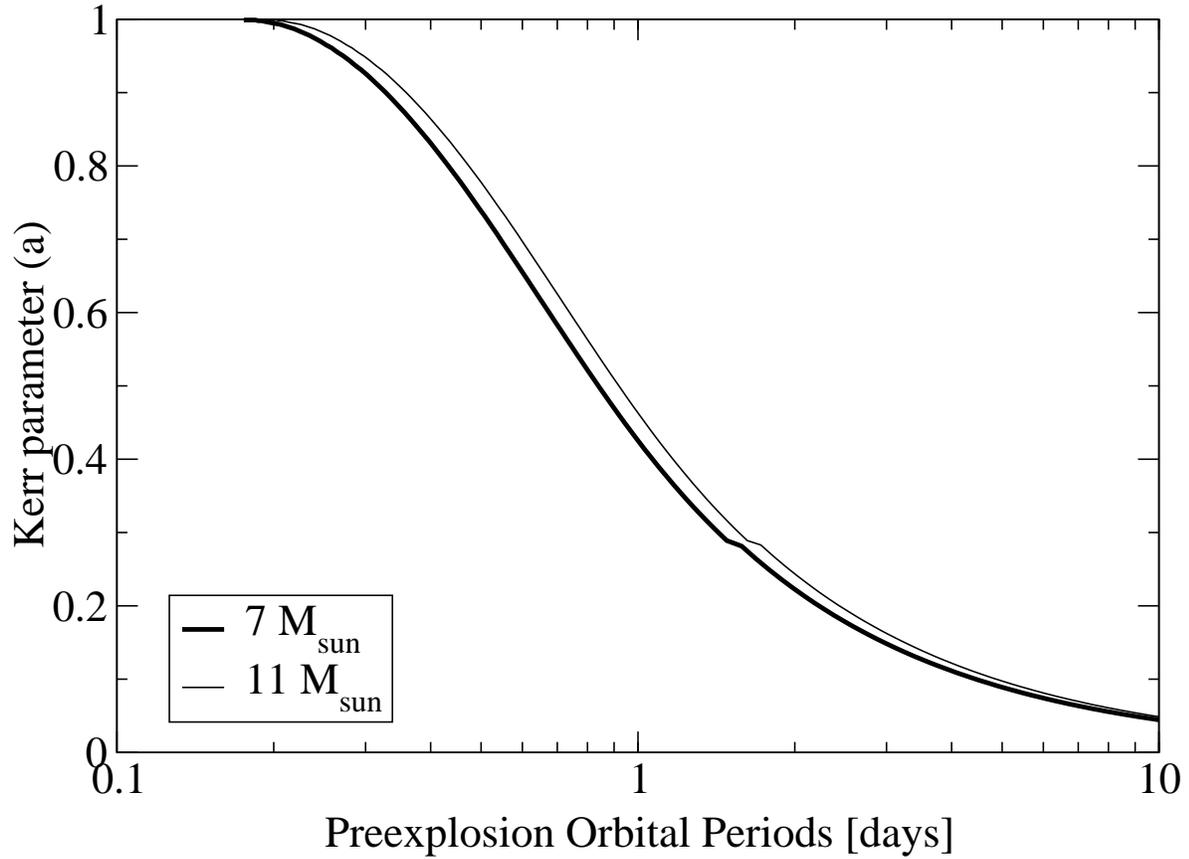}}
\caption{The Kerr
parameter of the black hole resulting from the collapse of a
helium star synchronous with the orbit, as a function of orbital
period (LBW). Note that the result depends very little on the mass
of the helium star, or on whether we use a simple polytrope or a
more sophisticated model. \label{FIG12} }
\end{figure}

From Kepler we have for the preexplosion period
\be
\frac{{\rm days}}{P_b}=\left(\frac{4.2\rsun}{a_f}\right)^{3/2}
\left(\frac{M_d+M_{\rm He}}{\msun}\right)^{1/2} \label{eq:Pb}
\ee
where
$M_d$ and $M_{\rm He}$ are the masses at the time of common envelope
evolution.  Given $P_b$ we can easily find the Kerr parameter
$a_{\star}$ from Fig. 12 of LBW, reproduced as Figure~\ref{FIG12}
here. In the case of Nova Sco, $P_b=1/4$day, $a_f=5.33\rsun$,
$M_{\rm He}=11\msun$ and $M_d=1.91\msun$ (LBW).

The big advantage that Case C mass transfer has is that it not
only produces an explosion with no hydrogen envelope, but
it produces a great deal of angular momentum, as quantified in the
Kerr parameter of the black hole, to power the GRB and Hypernova.
The angular momentum results from the tidal locking of the donor
and the He star, the latter falling into the black hole.
In the core, the helium is burned before common envelope evolution into carbon
and, rather quickly, oxygen.  The strong $\vec{B}$-field lines,
which at one end thread the disk of the black hole as it is formed
from the collapse inwards of the ionized metals, are frozen at the
other end in the metals and lock the disk tidally with these
metals which constitute what is left of the original helium star.
If we replace the helium star in the \citet{Mac99} paper
 by our He star then the formation
and spin up  of the black hole is as these authors described. Thus
we basically have a collapsar with high angular momentum that has
been spun up by tidal locking with the donor.
Note that there is no hydrogen envelope of the giant left, the hydrogen
having been expelled in common envelope evolution.

\section{GRBs and Hypernovae from Soft X-ray Transients With Evolved Companions}

In Table~\ref{tab-a} we list the black hole masses, and our
estimates of donor masses, all at the time of the end of common
envelope evolution when the tidal locking was established between
donor and helium star.  These came from LBW and from \citet{Mor07}.
 The Kerr parameters are changed into
energies using the Blandford-Znajek formulas \citep{Lee00}
 \be E_{BZ}=1.8\times10^{54}\epsilon_{\Omega}
f({a_{\star}}) \frac{M_{BH}}{\msun}{\rm ergs}\label{eq:BZ}\ee
where the efficiency $\epsilon_{\Omega}=\Omega_F/\Omega_H$ for
energy deposition in the (perturbative) fireball is taken to be
$1/2$ (for optimum impedance matching%\footnote{So we may
%overestimate the energies somewhat}
) and \be
f(a_{\star})=1-\sqrt{\frac{1}{2}(1+\sqrt{1-a_{\star}^2})}.
\label{eq:fa} \ee

We note that Cyg X$-$1 (1956$+$350y) probably went through a dark
explosion \citep{Mir03} meaning that at most, a
very low energy, one or two magnitudes less than in the case of
Nova Sco. The high Kerr parameter ($a_{\star}>0.98$) for GRS 1915$-$105
\citep{McC06} came chiefly from mass accretion
following the explosion in which the black hole was born, and,
therefore, had no influence on the GRB \citep{Mor07}.
The measured Kerr parameters are the present ones, and the
energies to produce the GRB and Hypernova should be subtracted
from our calculated ones.
%We find the available spin energies of Cyg X$-$1 and GRS 1915$-$105 to have been roughly the same as the explosion %energies of cosmological GRBs.

What we see from Table~\ref{tab-a} is that the transient sources
Nova Sco, Il Lupi, XTE J1550$-$564,   and GS 2023$+$338 clearly had
enough rotational energy to power both a GRB and Hypernova
explosion.  \citet{Bro00} in discussing these for Nova Sco
suggested that the energy was so great that the explosion disrupts
the accretion disk; this removes the magnetic fields anchored in
the disk and self-limits the energy the Blandford-Znajek mechanism
can deliver (see the appendix). In addition to the 7 sources in Table~\ref{tab-a}, LBW worked out the Kerr parameters of the 8 Galactic X-ray Transient sources with main sequence companions, all of which had $a_\star$'s of $0.6-0.8$ which correspond to spin energies of $430-600 \times 10^{51}$ ergs.

In \citet{Bro00} the GRB and hypernova explosion were reconstructed in all detail. The F-star donor in Nova Sco bore witness to the hypernova explosion through the $\alpha$-particle nuclei deposited on it. In particular, a large amount of Sulfur, which \citet{Nomo00} found typical of differentiating hypernovae from the more usual supernovae, was found. The Kerr parameter of 0.8 found by LBW for the preexplosion spin was, within uncertainties, the same as the post explosion Kerr parameters measured by \citet{Sha06}. The GRB was, of course, not recorded, but the rotational energy was tremendous so that the GRB was either just begun or the accretion disk was smashed immediately. The system velocity was worked out.  Almost all of the natal angular momentum energy is still in the system, as measured by \citet{Sha06}, meaning that very little was accepted for the explosion.

\section{Subluminous GRBs}

In \citet{Bro00} the population synthesis suggested a soft X-ray transient birth rate of $3\times 10^{-4}$ sources per year per galaxy, which with $10^5$ galaxies within 200 Mpc translates into 3750 Gpc$^{-3}$yr$^{-1}$. If we consider the beaming factor of $\sim10\%$, this is the same rate as the rate of subluminous sources investigated by \citet{Lia07}, estimated at $325^{+352}_{-177}$ Gpc$^{-3}$yr$^{-1}$. The latter are thought to have come from low-metallicity galaxies, but it is none the less interesting that the rate of hypernovae from soft X-ray transient sources is the same as that of the subluminous bursts, especially because we have shown that only a small part of the black hole spin energy in soft X-ray transient sources went into the explosion, so that they would tend to be subluminous.

The question of central engine for GRB060218 was tackled by the 119 astronomers who signed the 5 papers in Nature \citep{Cam06,Maz06,Pia06,Sod06,You06}. From the Supplementary Information of \citet{Maz06} one finds that the explosion 2006aj was Type I$_{bcd}$ in nature; i.e., in addition to no hydrogen lines, no helium nor carbon. The only place where this could occur was in a %n $18-20\msun$
black hole in which convective carbon burning ceases because the carbon abundance drops below 15\%: see Fig.~1 of \citet{Bro01}. This leaves no doubt but that the central engine was powered by a black hole, one of low mass.

Galactic GRBs (GRBs from galaxies with solar metallicity) must be subluminous, relatively little of their tremendous rotational energy being used up in the explosion. For the population of metal poor subluminous GRBs %are the same as for Galactic GRBs, but
one would expect
%their donors to be more massive, because of the low metallicity, so that although a greater fraction of their lower rotational energy would be used up in the explosion, they would still be subluminous. The host galaxy of GRB980425 does not seem to have metallicity much less than solar \citep{Sol05} and would be expected to not be much different from the Galactic GRBs. However, the recent measurements in M33 X-7 \citep{Oro07} in which $M_{\rm donor}=68.5\msun$, $M_{\rm BH}=15.4 \msun$ and orbital period of 3.45 days indicate that it had a rotational energy greater than $2\times 10^{52}$ ergs and was likely to have had about twice this energy unless the explosion was a ``smothered" one as proposed by \citet{Mac01} for GRB980425. The metallicity of M33 is $<10\%$ solar and the evolution should be similar to that of GRB980425, although in detail the evolution of low-metallicity binaries can be very complicated \citep{Can07}.
%%proofread Diane's:
their donors to be more massive %, since with low metallicity their evolution goes in the direction of population II stars, which are substantially larger than Galactic.
 because of their low metallicity.  Because of the more massive donors they will have less rotational energy, which may be all utilized in the explosions or, at least, will take larger to dismantle the disk.  Thus they would be of relatively long duration, but subluminous in the integrated energy in the explosion.

Recently an eclipsing binary M33 X-7 was discovered in a metal poor neighborhood ($\sim10\% $solar) by Orosz (2007).  This can be evolved like a more massive Cyg X-1, but with the advantage that one knows the donor to be $\sim80\msun$ at the time of explosion.  The Kerr parameter was $a_{\star}=0.12$ and the angular momentum energy of %$2 \times
$\sim10^{52}$ ergs was too little to both power the jet for a GRB and the hypernova, so the explosion was probably ``dark.''  Had the donor been less massive, according to our arguments, then with more energy the GRB and hypernova could have been powered.  We agree that the subluminous bursts come chiefly from metal poor galaxies \citep{Sta06}, giving the dynamical reason that they have low angular momentum energies because of larger donor masses.

%%There seem to be about as many
%subluminous GRBs and Hypernovae as cosmological ones \citep{Cam06,Maz06,Pia06,Sod06,You06}.
%We suggest that these came from the transient sources.

\section{Discussion}

We show that the rotational energy of black holes in soft X-ray transient sources is greatest when the donor in
the binary is of low mass.  In the case of large donor masses, the rotational energy in the black hole binary is lower.
%explosions in which the black hole is formed are of lower energy.

One can see that Nova Sco had a very high explosion energy from
the fact that its space velocity after the explosion is
$112\pm18$ km s$^{-1}$ as to compare with Cyg X$-$1 relative to
Cyg OB3 in the cluster of O-stars of $9\pm2$ km s$^{-1}$, which
is typical of the random velocities of stars in expanding
associations \citep{Mir03}.

The explanation of why the angular momentum energy is so high in
Nova Sco was given on p.176 of \citet{Bet03}:  {\it ``The
massive star will have evolved through its supergiant (He core
burning stage) before matter overflows its Roche lobe.  Then, by
that time, a main sequence companion must be at just the right
distance to receive the overflow; this means $a\sim1500\rsun$, the
Roche lobe of the massive star being at $\sim\frac{2}{3}a$.  Since
the binding energy of the envelope of the massive star goes as
$1/a$, this binding energy is very small, so that the envelope can
be removed by the drop in gravitational energy of an $\sim1\msun$
main sequence star as it moves inwards in common envelope
evolution with the massive star from $\sim1500\rsun$ to the much
smaller Roche lobe of the He star which results when the H
envelope is removed from the massive star. In this way one could
understand why all of the main sequence companions of the black
holes in the transient sources were of nearly the same low masses,
$(0.5-1)\msun$."}
  For the companions with masses $10\msun$, the
necessary drop in gravitational energy is only $1/10$ that of the
$1\msun$ companion, so the final $a_f$ can be an order of
magnitude greater.  The result is an order of magnitude lower
rotational energy.

From the above explanation we see that the ultrahigh
rotational energies in soft X-ray transient sources are a result of the low donor masses.  The
rotational energy drops roughly inversely with mass so we would
expect it to be an order of magnitude less for %Population II and
%III stars,
stars of low metallicity whose masses are roughly an order of magnitude greater
than stars in our Galaxy.  Thus, cosmological GRBs will not have so much
rotational energy as to dismantle the disk, and may be able to
furnish their rotational energy to the GRB and Hypernova.  At
least, now that we understand why Galactic GRBs  are so energetic,
we can offer reasons why the cosmological GRBs %will be less so.
have lower energy, but may be able to use up more of it in the explosion.

Measurement of the Kerr parameter for XTE J1550-564 (J. McClintock et al., Smithsonian-Harvard coalition, in progress) will enable us to say how much of the natal $\sim300$Bethes was used up in the explosion.

%In conclusion, we believe the \citet{Bro00} theory of GRBs
%and Hypernova explosions to explain those of the Galactic
%transient sources and suggest that it may also explain cosmological
%GRBs.

\section{Summary}

In summary, the essential points of our paper are that the Woosley Collapsar model can be
obtained from our Case C mass transfer, but with the black hole having any desired angular momentum, by making choice of the donor mass. Because the helium is burned preceding the explosion in Case C mass transfer, the ashes of the central helium, carbon and oxygen, fall first into the black hole and ensure the tidal locking through the strong B-field lines which are frozen in the ionized metals.

Our results for the LBW calculation of Kerr parameters have been confirmed by the Smithsonian-Harvard group. Given the Kerr parameters, we can make quantitative calculation of the spin energy of black hole. We give predictions for the Kerr parameters of 12 Galactic black hole sources which have not yet been measured.

We note that the rotational energy of M33 X-7 was %approximately
%somewhat
lower than that of cosmological GRBs and suggest that these originate from %Population II
low metallicity donors of somewhat less mass than that of M33 X-7.  Our suggestion that XTE J1550-564 should have the angular momentum energy in its explosion, which is as large as that of cosmological GRBs, should soon be tested by the measurement of the Kerr parameter.

%\acknowledgements

\section*{Acknowledgments}

We would like to thank Jeff McClintock for many useful discussions.
G.E.B. was supported by the US Department of Energy under Grant No. DE-FG02-88ER40388.
C.H.L. was supported by Grant No. R01-2005-000-10334-0 (2005)
from the Basic Research Program of the Korea Science \& Engineering Foundation.

%--------------------------------------------------------------------
\appendix

\section*{Dismantling the Accretion Disk
            by High Energy Input}\label{appendix}

Knowing the Kerr parameters we can make quantitative estimates of energy. The amount of energy poured into the accretion disk of the black
hole, and, therefore, also pressure is almost unfathomable., the
$5\times10^{53}$ergs being $500$ times the energy of a strong
supernova explosion, the latter being spread over a much larger
volume than that of the accretion disk.  Near the horizon of the
black hole, the physical situation might become quite complicated \citep{Tho86}.
Field-line reconstruction might be common and lead to serious
breakdowns in the freezing of the field to the plasma; and the
field on the black hole sometimes might become so strong as to
push its back off the black hole and into the disk
(Rayleigh-Taylor Instability) concentrating the energy even more.
During the instability the magnetic field lines will be
distributed randomly in ``globs", the large ones having eaten the
small ones.  It seems reasonable that the Blandford-Znajek
mechanism is dismantled.  Later, however, conservation laws demand
that the angular momentum not used up in the GRB and hypernova
explosion be reconstituted in the Kerr parameter of the black
hole.

%%%%%%%%%%%%%%%%%%%%%%%%%%%%%%%%%%%%%%%%%%%%%%%%%%%%%%%%%%%%%%%%%%%%%%%%%%%%%%%%

%------------------------------------------------------
\clearpage

\begin{table}[h]
\begin{center}
\begin{tabular}{|c|c|c|c|c|}
\hline
 Name           &$M_{BH}$ &  $M_{d}$ &$a_{\star}$&   $E_{\rm BZ}$    \\
                &[$\msun$]& [$\msun$]&           &    [Bethes ]      \\
\hline
GRO J1655$-$40  & $\sim5$ &  1$-$2   &   $0.8$   & $\sim 430$ \\
4U 1543$-$47    & $\sim5$ &  1$-$2   &   $0.8$   & $\sim 430$ \\
\hline
XTE J1550$-$564 &$\sim10$ &  1$-$2   &   $0.5$   & $\sim 300$ \\
GS 2023$+$338   &$\sim10$ &  1$-$2   &   $0.5$   & $\sim 300$ \\
\hline
XTE J1819$-$254 &  6$-$7  & $\sim10$ &    0.2       & $10\sim 12$ \\
GRS 1915$+$105  &  6$-$7  & $\sim10$ &  0.2 ($>0.98^\dagger$)  & $10\sim 12$ \\
Cyg X$-$1       &  6$-$7  &$\gtrsim30$& 0.15         & $5\sim 6$ \\
\hline
\end{tabular}
\end{center}
\caption{Parameters at the time of black hole formation.
$E_{\rm BZ}$ is the rotational energy which can be extracted via Blandford-Znajek mechanism with optimal efficiency $\epsilon_\Omega=1/2$ in Eq.~(\ref{eq:BZ}), except for low $a_{\star}$s, $a_{\star}=0.2$ and $0.15$ where the efficiencies are lower, give $\epsilon_\Omega=0.37$ and $0.33$ as calculated in Appendix 2 of \cite{Bro00}.
$^\dagger$ Kerr parameter is the present one.}\label{tab-a}
\end{table}

%\clearpage

\begin{thebibliography}{99}
%--------------------------------------------------------------------

\bibitem[Bethe et al.(2003)]{Bet03}
Bethe, H.A., Brown, G.E., and Lee, C.-H. 2003, {\it Formation and
Evolution of Black Holes in the Galaxy : Selected Papers with
Commentary}, World Scientific Series in 20th Century Physics Vol.
33 (World Scientific, 2003).

%\bibitem[Blaauw(1961)]{Bla61}
%Blaauw, A. 1961, {\it BAN} 15, 265.
%
%\bibitem[Boersma(1961)]{Boe61}
%Boersma, J. 1961, {\it BAN} 15, 291.

\bibitem[Brown et al.(2001)]{Bro01}
Brown, G.E., Heger, A., Langer, N., Lee, C.-H., Wellstein, W., and Bethe, H.A. 2001, {\it New Astronomy}, 6, 457.

\bibitem[Brown et al.(2000)]{Bro00}
Brown, G.E., Lee, C.-H., Wijers, R.A.M.J. Lee, H.K., Israelian, G.
and Bethe, H.A. 2000, {\it New Astronomy} 5, 191.

\bibitem[Campana et al.(2006)]{Cam06}
Campana, S. et al. 2006, {\it Nature} {\bf 442}, 1008.

%\bibitem[Heger et al.(2003)]{Heg03}
%Heger, A., Woosley, S.E., Langer, N. and Spruit, H.C. 2003, {\it Proc.
%IAU Symposium} No. 215, 2003 IAU.

\bibitem[Cantiello et al.(2007)]{Can07}
Cantiello, M., Yoon, S.-C., Langer, N., and Livio, M. 2007, {\it A\&A} 465, L29.

\bibitem[Lee et al.(2000)]{Lee00}
Lee, H.K., Wijers, R.A.M.J., Brown, G.E. 2000, {\it Physics Reports}
325, 83.

\bibitem[Lee et al.(2002)]{Lee02}
Lee, C.-H., Brown, G.E., and Wijers, R.A.M.J. 2002, {\it ApJ},
575, 996 (LBW).

\bibitem[Liang et al.(2007)]{Lia07}
Liang, E., Zhang, B., Virgili, F., and Dai, Z.G. 2007, {\it ApJ}, 662, 1111.

\bibitem[MacFadyen \& Woosley(1999)]{Mac99}
MacFadyen, A.I. and Woosley, S.E. 1999 {\it ApJ} 524, 262.

\bibitem[MacFadyen et al.(2001)]{Mac01}
MacFadyen, A.I., Woosley, S.E., and Heger, A. 2001, {\it ApJ}, 550,410.

\bibitem[Mazzali et al.(2006)]{Maz06}
Mazzali, P.A. et al. 2006, {\it Nature} {\bf 442}, 1018.


\bibitem[McClintock et al.(2006)]{McC06}
McClintock, J.E., Shafee, R., Narayan, R., Remillard, R.A., Davis,
S.W., and Li, L.-X. 2006, {\it ApJ}, 652, 518.


\bibitem[Mirabel \& Rodrigues(2003)]{Mir03}
Mirabel, I.F., Rodrigues, I.  2003, {\it Science}, 300, 1119.

\bibitem[Moreno M\'endez et al.(2007)]{Mor07}
Moreno M\'endez, E., Brown, G.E., Lee, C.-H. and Walter, F. 2007,
in progress.

\bibitem[Nomoto et al.(2000)]{Nomo00}
Nomoto et al., The Greatest Explosions since the Big Bang: Supernovae and Gamma-Ray Bursts, CUP:Cambridge; astro-ph/0003077.

\bibitem[Orosz et al.(2007)]{Oro07}
Orosz, J. et al. 2007, {\it Nature}, {\bf 449}, 872.

\bibitem[Pian et al.(2006)]{Pia06}
Pian, E. et al. 2006, {\it Nature} {\bf 442}, 1011.


\bibitem[Shafee et al.(2006)]{Sha06}
Shafee, R., McClintock, J.E., Narayan, R., Davis, S.W., Li, L.,
and Remillard, R.A. 2006, {\it ApJ}, 636, L113.

\bibitem[Soderberg et al.(2006)]{Sod06}
Soderberg, A.M. et al. 2006, {\it Nature} {\bf 442}, 1014.

\bibitem[Sollerman et al.(2005)]{Sol05}
Sollerman, J., \"{O}stlin, G., Fynbo, J.P.U., Hjorth, J., Fruchter, A., and Pedersen, K. 2005, {\it New Astronomy}, 11, 103.

\bibitem[Stanek et al.(2006)]{Sta06}
Stanek, K.Z. et al. 2006, {\it Acta Astronomica}, 56, 333.

\bibitem[Thorne et al.(1986)]{Tho86}
Thorne, K.S., Price, R.H. and Macdonald, A., {\it Black Holes: The
Membrane Paradigm} New Haven and London, 1986, p.133.

%
%\bibitem[van den Heuvel \& Yoon(2007)]{vdH07}
%van den Heuvel, E.P.J., and Yoon, S.-C. 2007, astro-ph/0704.0659.
%

\bibitem[Young(2006)]{You06}
Young, T.R. 2006, {\it Nature} {\bf 442}, 992.


\end{thebibliography}
\end{document}